\documentclass{article}
\usepackage{spconf,amsmath,graphicx}

\usepackage{booktabs}
\usepackage{tabularx}
\usepackage{makecell}
\usepackage{multirow}
\usepackage{hyperref}
\usepackage{booktabs}
\usepackage{tabularx}
\usepackage{makecell}
\usepackage{array}

\newif\ifblind
\blindfalse   % 카메라레디(저자 포함) 만들 때

\usepackage{textcomp}

\title{Repurposing Image Diffusion Models for Training-Free Music Style Transfer on Mel-spectrograms}

\ifblind
\name{}      % keep empty for double-blind
\address{}   % keep empty for double-blind

\else
\name{%
\begin{tabular}{c}
Heehwan Wang$^{1,*}$\thanks{* Equal contribution.},\;
Joonwoo Kwon$^{2,*}$,\;
Sooyoung Kim$^{3,*,\dagger}$\thanks{$\dagger$ This research was conducted at the authors’ previous institution, SNU.}, \\
\textit{Jungwoo Seo}$^{1}$,\;
\textit{Shinjae Yoo}$^{4,\ddagger}$,\;
\textit{Yuewei Lin}$^{4,\ddagger}$,\;
\textit{Jiook Cha}$^{1,\ddagger}$\thanks{$\ddagger$ Co-corresponding authors. \\ This work is supported by IITP grant (MSIT, No.~RS-2021-II211343, AI Graduate School Program, SNU); NRF grants funded by MSIT (Nos.~2021R1C1C1006503, RS-2023-00266787/00265406, RS-2024-00421268/00342301/00435727, RS-2025-25457239, RS-2021-NR061370, NRF-2021M3E5D2A01022515, NRF-2021S1A3A2A02090597); and SNU Researchers Programs (Nos.~200-20250071/49/116/115/113, 0670-20250039, 200-20260009). This work was also supported by the U.S.~DOE, under award DE-SC-0012704. This research used resources of the National Energy Research Scientific Computing Center, a DOE Office of Science User Facility using NERSC award ALCC-ERCAP0030659.}
\end{tabular}
}
\address{$^1$ Seoul National University \quad
         $^2$ Michigan State University \quad
         $^3$ Rutgers University \\
         $^4$ Brookhaven National Laboratory}
\fi

\begin{document}
%\ninept
%
\maketitle
\begin{abstract}

% Music style transfer enables personalized creation by blending source structure with reference style. However, existing zero-shot approaches often fail to capture fine-grained audio nuances, relying on coarse text descriptions or requiring expensive task-specific training. We present \textbf{Stylus}, a \textit{training-free} framework that repurposes pretrained image diffusion models (e.g., Stable Diffusion) for music style transfer in the Mel-spectrogram domain. By treating audio as structured time–frequency images, Stylus manipulates self-attention—injecting style keys and values while preserving source queries—to maintain musical structure. To ensure high fidelity, we introduce a \textit{phase-preserving reconstruction} strategy that mitigates spectrogram inversion artifacts, alongside a classifier-free-guidance-inspired control for adjustable stylization. Extensive evaluations show that Stylus outperforms state-of-the-art baselines, achieving 34.1$\%$ higher content preservation and 25.7$\%$ better perceptual quality. Ultimately, this work bridges image diffusion, attention manipulation, and audio signal processing, highlighting how generic image priors can be effectively leveraged for the training-free transformation of structured Mel-spectrogram images. Code and materials will be made publicly available upon acceptance.

Music style transfer blends source structure with reference style to enable personalized music creation. However, existing zero-shot methods often struggle to capture fine-grained audio nuances, relying on coarse text descriptions or requiring expensive task-specific training. We propose \textbf{Stylus}, a training-free framework that repurposes pretrained image diffusion models for music style transfer in the Mel-spectrogram domain. By treating audio as structured time-frequency images, Stylus manipulates self-attention by injecting style keys and values while preserving source structural queries. To ensure high fidelity, we introduce a phase-preserving reconstruction strategy to mitigate spectrogram inversion artifacts, alongside a classifier-free-guidance-inspired control for adjustable stylization. Extensive evaluations including 2,925 human ratings demonstrate that Stylus outperforms state-of-the-art baselines, achieving $34.1\%$ higher content preservation and $25.7\%$ better perceptual quality. Our work validates that generic image priors can be effectively leveraged for the training-free transformation of structured Mel-spectrograms. Code and materials are available at \url{https://github.com/Sooyyoungg/Stylus.git}.
\end{abstract}

\begin{keywords}
music style transfer, training-free, image diffusion models, phase-aware synthesis, Mel-spectrogram
\end{keywords}
\section{Introduction}
\label{sec:intro}

Music style transfer blends the structural elements of one musical piece with the stylistic attributes of another, enabling expressive and personalized music creation. Despite remarkable progress in generative music models, existing approaches often struggle to support fine-grained style manipulation while preserving musical structure. In practice, stylistic attributes such as timbre and temporal texture are perceptually salient yet difficult to specify precisely using text descriptions alone.

Many recent diffusion-based models \cite{huang2023noise2music, copet2023simple}, including MusicLM \cite{agostinelli2023musiclm}, Riffusion \cite{Forsgren_Martiros_2022}, and Make-An-Audio \cite{huang2023make}, therefore rely on text prompts to describe style. While text-based conditioning offers intuitive control, it often struggles to capture the nuanced characteristics present in audio exemplars, resulting in transferred styles that are coarse or only loosely aligned with the reference.

As an alternative to text-based conditioning, audio-to-audio diffusion methods (e.g., MusicTI \cite{li2024music}) move closer to exemplar-based style transfer by conditioning directly on audio signals. However, these approaches typically require task-specific training or fine-tuning, such as learning dedicated style encoders or adapting model parameters to predefined style categories. This reliance on additional training limits scalability and makes zero-shot deployment difficult, particularly when users provide reference tracks with novel or unlabeled timbral characteristics.

These challenges stem fundamentally in problem formulation rather than implementation choices. Most existing diffusion-based music models are developed with synthesis and generation as their primary objective and are therefore optimized to encode domain-specific musical regularities, such as harmonic organization and long-range temporal dependencies. While effective for generation, this formulation falls short in contexts where style must be specified through concrete audio exemplars without additional training. In such cases, exemplar-based style transfer in a \emph{training-free} setting imposes a different requirement: the model must selectively modify stylistic attributes while preserving content, using a fixed pretrained representation without further adaptation.

To address this zero-shot setting, we fundamentally revisit the representation of music. Instead of treating music as a waveform generation task, we interpret Mel-spectrograms as structured time-frequency images, encoding musical content and style as 2D patterns such as harmonic textures and temporal envelopes. While audio-native diffusion models are trained to enforce domain-specific regularities such as harmonic structure and waveform realism, which can limit flexibility for style manipulation, image diffusion models instead learn generic spatial statistics. When applied to Mel-spectrograms, this generic spatial prior enables the flexible manipulation of local time-frequency patterns without being constrained by learned musical rules, making it uniquely suited for training-free, exemplar-based style transfer.

Motivated by this, we introduce \textbf{Stylus}, a framework that repurposes a pretrained image diffusion model \cite{rombach2022high} to perform \textit{training-free} music style transfer directly in the Mel-spectrogram domain. By operating on spectrograms as image-like representations and manipulating self-attention layers, Stylus enables exemplar-based style transfer without requiring any task-specific training or fine-tuning.

Specifically, Stylus injects style key-value features into self-attention while preserving content queries, enabling controlled modification of stylistic attributes while maintaining musical structure. To improve synthesis fidelity after spectrogram-level manipulation, we introduce a \textit{phase-preserving reconstruction} strategy that reduces artifacts commonly introduced by iterative phase estimation methods, such as Griffin--Lim~\cite{griffin1984signal}. In addition, we adopt classifier-free-guidance-inspired control to enable continuous adjustment of stylization strength and multi-style blending at inference. This work bridges audio signal processing with generative computer vision, validating that generic spatial priors can be robustly repurposed for zero-shot acoustic transformation.

Our contributions are summarized as follows:
\begin{itemize}
  \item We propose a \textit{training-free} diffusion-based framework, \textbf{Stylus}, that formulates music style transfer as attention-based manipulation of Mel-spectrogram images.
  \item We introduce \textbf{controllable attention manipulation} and \textbf{phase-preserving reconstruction} to jointly improve flexibility and perceptual fidelity without additional training.
  \item We provide extensive experimental evaluations against state-of-the-art baselines, demonstrating that Stylus achieves substantially improved content preservation and perceptual quality in a zero-shot setting.
\end{itemize}

%%%%%%%%

%-----------------------------------------------------------------------
\section{Related Work} \label{sec:related_work}

\textbf{Music Style Transfer.} Music style transfer seeks to recombine structural (melody, rhythm) and stylistic (timbre, texture) elements~\cite{dai2018music}. Early work emphasized melody-preserving timbre transfer with WaveNet autoencoders~\cite{engel2017neural} and CNNs~\cite{grinstein2018audio}, later extending to genre transfer via adversarial frameworks such as CycleGAN~\cite{huang2018timbretron}. Recent large language and diffusion-based models~\cite{agostinelli2023musiclm,copet2024simple,huang2023noise2music,li2024music,kwon2025revisiting} enable controllable synthesis via text or auxiliary conditions, but typically demand extensive training and rely on Griffin--Lim~\cite{griffin1984signal} for reconstruction, which introduces artifacts after transfer. In contrast, Stylus performs structure-aware style transfer directly on Mel-spectrograms and reuses the content phase, substantially reducing artifacts and achieving high fidelity without any additional training.

\textbf{Image Diffusion Models.} Diffusion models have surpassed GANs in generative fidelity owing to training stability and scalability~\cite{dhariwal2021diffusion,ho2020denoising}. Non-Markovian samplers like DDIM~\cite{song2020denoising} accelerate inference, and Stable Diffusion (SD)~\cite{rombach2022high} pioneered efficient high-quality synthesis in a compressed latent space. Adaptation to music, however, remains hindered by data scarcity and prohibitive training costs, which we bypass by repurposing SD for training-free style transfer.

%-----------------------------------------------------------------------

\section{Method}
\label{sec:methods}
\textbf{Stylus} performs \textit{training-free} exemplar-based music style transfer by manipulating intermediate representations in a fixed pretrained image diffusion model (e.g., Stable Diffusion~\cite{rombach2022high}). Audio waveforms are converted into Mel-spectrogram images via Short Time Fourier Transform (STFT) and a Mel filter bank, normalized to $[0,1]$, and projected into the latent space through DDIM inversion~\cite{song2020denoising}.

\subsection{Attention-based Style Manipulation}
Image diffusion models capture generic spatial statistics over two-dimensional signals. When Mel-spectrograms are treated under this spatial prior, self-attention layers operate on time--frequency patterns as visual textures, enabling timbral style manipulation without being constrained by the strict harmonic dependencies inherent in audio-native models. This property makes attention manipulation a natural mechanism for exemplar-based style transfer in the spectrogram domain. By leveraging the inductive bias that self-attention queries ($Q$) anchor structural geometry while keys ($K$) and values ($V$) encode stylistic texture, we can inject novel timbres without altering the original musical layout.

Following \cite{chung2024style}, we adapt self-attention features via cross-attention, conditioning on the style spectrogram $I^{s}$. Specifically, during the generation process, we replace the \textit{key} and \textit{value} features of the content music's Mel-spectrogram with those from the style music's Mel-spectrogram. To achieve this, we first obtain the latent representations for both the content and style Mel-spectrograms, and then capture the self-attention features of the style Mel-spectrogram through DDIM inversion \cite{song2020denoising}. For predefined timesteps $t = {0, ..., T}$, we invert the style and content Mel-spectrograms, denoted as $z_{0}^{m_c}$ and $z_{0}^{m_s}$, from the image space ($t=0$) to Gaussian noise ($t=T$). During this process, we also gather the \textit{query} features of the content Mel-spectrogram ($Q_{t}^{m_c}$) and the \textit{key} and \textit{value} features of the style Mel-spectrogram ($K_{t}^{m_s}$, $V_{t}^{m_s}$) at each timestep. We initialize the stylized latent noise $z_{T}^{m_o}$ by combining the style $z_{T}^{m_s}$ and content latent representation $z_{T}^{m_c}$ through AdaIN~\cite{huang2017arbitrary}. The style transfer is carried out by replacing the original \textit{key} $K_{t}^{m_o}$ and \textit{value} $V_{t}^{m_o}$ in the self-attention layer with the \textit{key} $K_{t}^{m_s}$ and \textit{value} $V_{t}^{m_s}$ derived from the style Mel-spectrogram, during the reverse process of generating the stylized output latent $z_{0}^{m_o}$. To preserve the content structure and avoid artifacts, we implement a query preservation technique as follows:

\begin{equation}
\begin{gathered}
\bar{Q}_{t}^{m_o} = \gamma \times Q_{t}^{m_c} + (1 - \gamma) \times Q_{t}^{m_o}, \\
\phi_{out}^{m_o} = Attn({\bar{Q}_{t}^{m_o}}, K_{t}^{m_s}, V_{t}^{m_s}),
\end{gathered}     
\end{equation}
where $\gamma$ is a hyperparameter. These operations are applied to the later layers of the decoder (layers 7–12 in the SD model) that focus on capturing local texture features.

\begin{figure*}[!t]
    \centering
    \includegraphics[width=\textwidth]{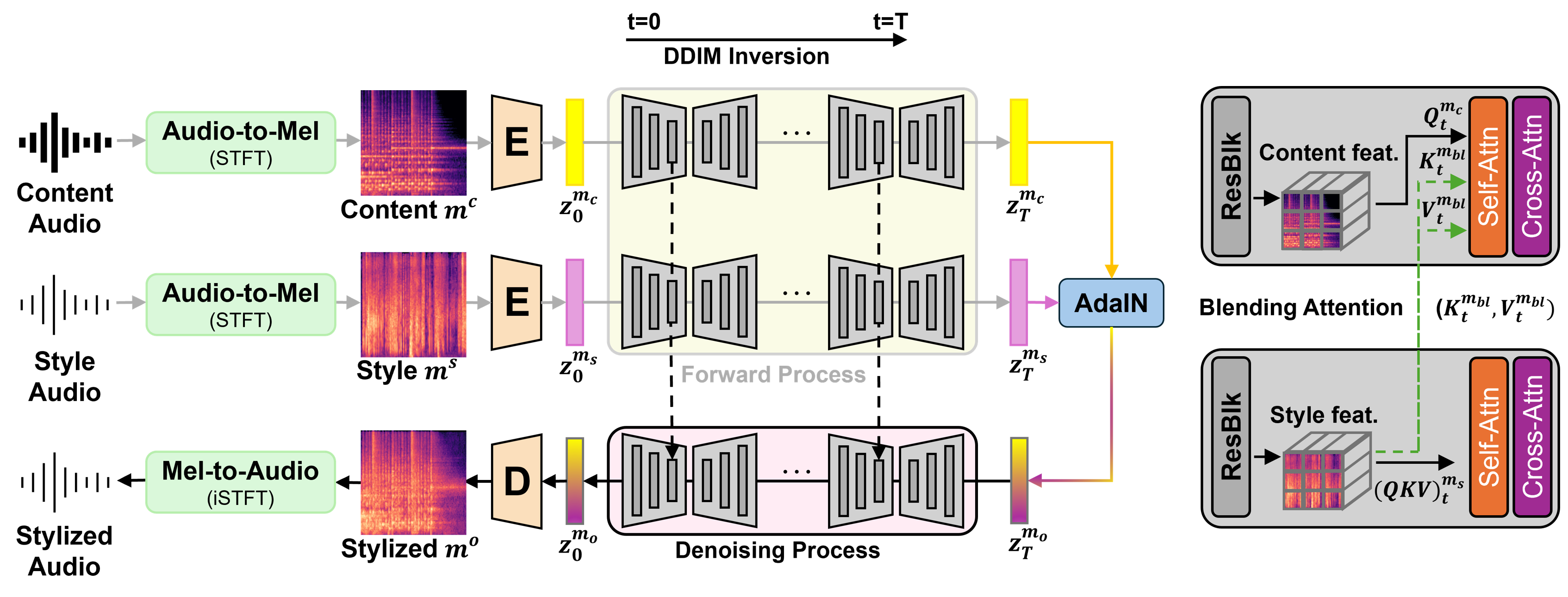}
    \caption{\textbf{Overall architecture} of \textbf{Stylus}. 
    Stylus adapts a pretrained diffusion model for music style transfer in the Mel-spectrogram domain. Key/value features in self-attention are swapped between content and style, while classifier-free-guidance-inspired blending provides controllable stylization. Phase-preserving reconstruction reuses the content phase to reduce artifacts. Note that our model requires \textbf{no additional training}.}
    \label{fig:fig1}
\end{figure*}

\subsection{Phase-Preserving Reconstruction}
Most prior music style transfer methods \cite{li2024music, huang2023noise2music} reconstruct waveforms from modified magnitudes using iterative phase estimation techniques such as Griffin--Lim~\cite{griffin1984signal}. While widely adopted, these methods often introduce audible metallic artifacts, including temporal smearing and unstable transients, which can significantly degrade perceptual quality, as demonstrated in the supplementary audio samples.

To mitigate these reconstruction artifacts without requiring additional training, Stylus purposefully employs a phase-preserving strategy by reusing the original content phase. In the context of exemplar-based, \emph{training-free} style transfer, we optimize for acoustic stability to ensure high perceptual fidelity. Since the stylized outputs lack a ground-truth phase, unconstrained phase estimation from modified magnitudes often yields perceptually implausible results. By explicitly locking the phase, our design circumvents these artifacts while effectively transferring stylistic attributes through the magnitude domain.

%To mitigate these reconstruction artifacts in a training-free setting, Stylus adopts a phase-preserving strategy that reuses the original content phase. This is indeed a restrictive design choice, as it limits the transfer of stylistic attributes encoded in spectral phase. However, in the context of exemplar-based, \emph{training-free} music style transfer, this choice reflects an explicit trade-off between stylistic completeness and perceptual fidelity. Since the stylized outputs are not reconstructions of an existing recording, no ground-truth phase is available, and unconstrained phase estimation from modified magnitudes often leads to perceptually implausible results.

Consistent with this observation, our experiments show that phase-preserving reconstruction improves both subjective perceptual quality and content preservation compared to Griffin--Lim--based reconstruction (Table~\ref{tab:ablation1}), supporting its effectiveness as a pragmatic and reliable reconstruction strategy for training-free music style transfer.

\begin{figure}[!t]
    \centering
    \includegraphics[width=0.48\textwidth]{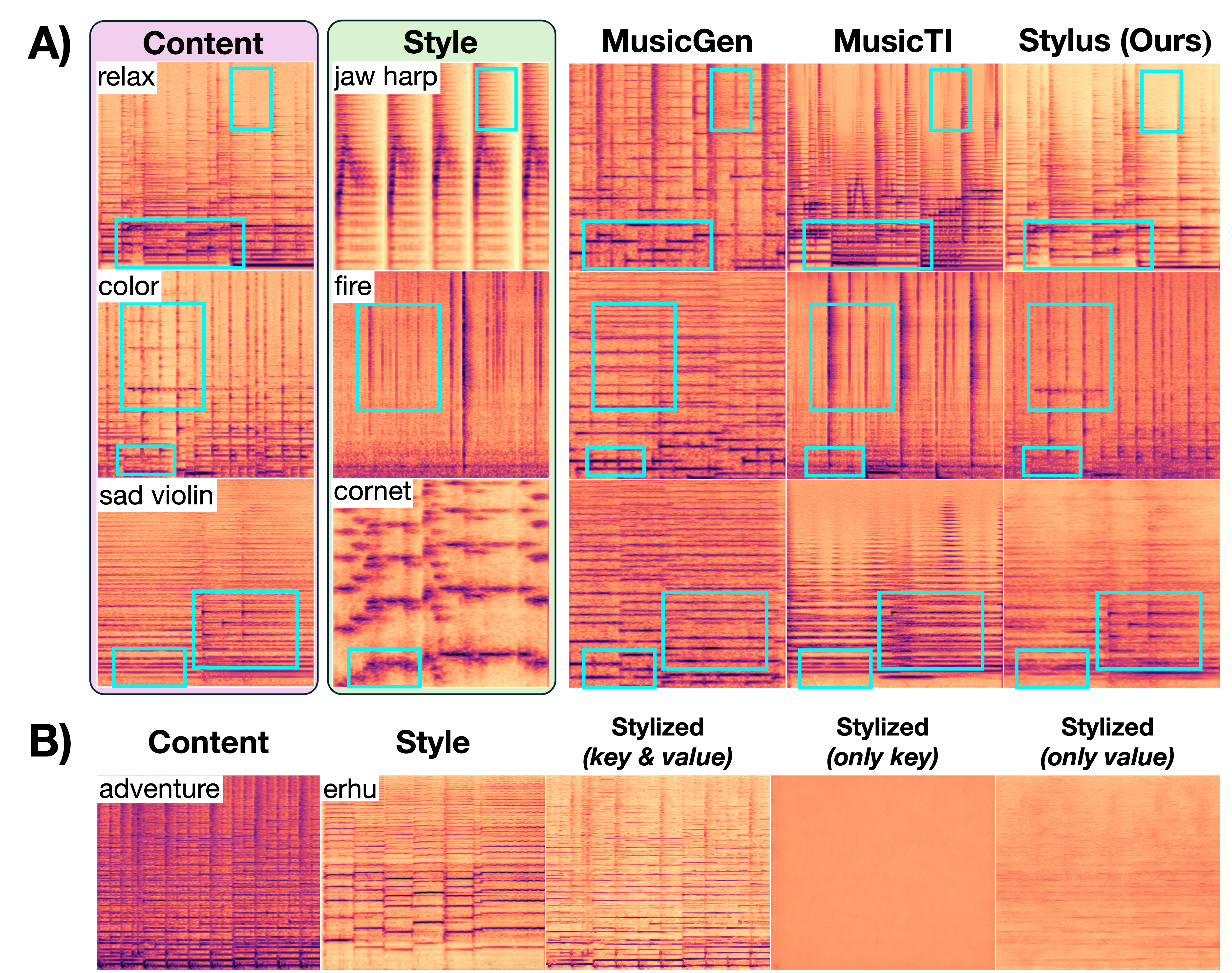}
    \caption{\textbf{Qualitative comparisons} A) Stylus effectively transfers the target style while preserving content. B) Ablation shows that removing \textit{key} or \textit{value} leads to blurry and degraded outputs, confirming their importance for stylization.}
    \label{fig:fig2}
\end{figure}

\subsection{Controlling Stylistic Intensity via Guidance Scaling} 
A key limitation of prior style-transfer approaches is that direct key–value replacement in self-attention layers enforces a binary choice between content and style, offering little flexibility. This rigid formulation often distorts continuous musical structures such as pitch contours, rhythmic onsets, or harmonic progressions.  

To address this, Stylus introduces a training-free interpolation strategy inspired by Classifier-Free Guidance (CFG)~\cite{ho2022classifier}. Instead of swapping features, we compute attention outputs for both content and style independently:
\begin{equation}
\phi_{\text{content}} = \text{Attn}(Q, K_c, V_c), \quad 
\phi_{\text{style}} = \text{Attn}(Q, K_s, V_s),
\end{equation}
and blend them as
\begin{equation}
\phi_{\text{bl}} = \phi_{\text{content}} + \alpha \cdot (\phi_{\text{style}}-\phi_{\text{content}}),
\end{equation}
where $\alpha \in [0,1]$ acts as a \textit{style guidance scale}. This formulation enables smooth control over stylization intensity: smaller $\alpha$ values preserve structural fidelity, while larger values emphasize stylistic characteristics.  

\textbf{Implementation Details.} We evaluate our method on the MusicTI dataset~\cite{li2024music}, comprising 253 five-second audio clips (74 style and 179 content). Given our \textit{training-free} pipeline, all samples are reserved exclusively for inference. We utilize pretrained Stable Diffusion v1.5~\cite{rombach2022high} with DDIM sampling~\cite{song2020denoising} over 50 timesteps ($t \in \{1, \dots, 50\}$). The parameters $\alpha=0.9$ and $\gamma=0.75$ function as inference-time controls, selected through a qualitative pilot study on independent samples rather than test-set optimization. This protocol ensures our findings demonstrate robust generalization without overfitting, upholding experimental rigor. All experiments were conducted on a single NVIDIA A100 (40G) GPU.

\section{Experimental Results}
\label{sec:results}
We benchmark our model against SOTA baselines using official implementations with default configurations. For MusicTI \cite{li2024music}, we trained the style encoder following the authors' protocol; for MusicGen \cite{copet2024simple}, content audio served as the melody guide with text-based style descriptions. Unlike baselines limited by text or specific encoders, our model accepts arbitrary audio exemplars, allowing us to evaluate all possible content-style combinations. %This produced 13,246 outputs, representing a fourfold increase over the 3,222 samples utilized by MusicGen and MusicTI.

%==================
%\textbf{Qualitative Comparisons.} As illustrated in Figure~\ref{fig:fig2}, Stylus generates high-fidelity stylized outputs across diverse instruments and musical elements. Visual inspection of the Mel-spectrograms confirms that our approach maintains the harmonic and rhythmic integrity of the content music significantly better than baseline models, while faithfully incorporating the target style. For example, when applying a \textit{jaw harp} style to the \textit{relaxing} track, Stylus successfully replicates the characteristic periodic decay and plucky texture by modulating high-frequency energy. Similarly, it seamlessly integrates the ambient, noisy texture of \textit{fire} sounds into the \textit{color} track without compromising the underlying musical structure.  Crucially, our model demonstrates the ability to perform both timbral and temporal manipulation. It effectively substitutes the clean, defined texture of a \textit{violin} with the breathy, indistinct quality of a \textit{cornet}. Moreover, it captures the temporal articulation of an \textit{accordion}, transforming the violin's short, detached notes into long, sustained phrases. These results stand in strict contrast to competing methods that often distort the musical backbone, highlighting our model's superior capability to disentangle and reapply stylistic attributes.

\textbf{Qualitative Comparisons.} As shown in Figure~\ref{fig:fig2}, Stylus preserves structural integrity significantly better than baselines while faithfully transferring style. It successfully injects complex textures, such as the \textit{jaw harp}'s periodic decay and \textit{fire} ambience, without disrupting the source melody. Beyond texture, Stylus adapts temporal dynamics by transforming a \textit{violin}'s detached notes into the sustained phrasing of an \textit{accordion} or the breathy timbre of a \textit{cornet}. These results highlight our model's capability to disentangle content structure from stylistic attributes without the distortion observed in competing methods.
%%%%%%%%%%%%%%%%%%%%%%%%%%%%%%%%%%%%%%%%%%%%%%%%%%%%%%%%%%%%%%%%%%%%%

\textbf{Quantitative Comparisons.} Following established protocols \cite{cifka2021self, li2024music}, we evaluate content preservation and style fit using CLAP scores, with style fit disaggregated into musical instrument (M\_Style) and sound effect (S\_Style) categories, depending on the type of musical sound. Our model's training-free design enables a uniquely comprehensive validation; we test on all 13,246 samples—a nearly four-fold increase in scale over prior work—contextualized by reference scores as described in MusicTI and MusicGen. This large-scale evaluation (Table ~\ref{tab:quantitative}) confirms our model's outperformance in content preservation and style fit for sound effects. Our method significantly outperforms MusicTI, achieving a 34.1\% higher score in content preservation and an 11.9\% higher score in sound effects style fit. These results highlight the effectiveness of our attention-based design and phase-preserving reconstruction strategy.

%%%%%%%%%%%%%% User Study%%%%%%%%%%%%%%%%%%%%%

% -----Quantitative v3-----
\begin{table}[ht]
\renewcommand*{\arraystretch}{1.2}
\centering
\caption{\textbf{Quantitative Comparison} with various state-of-the-art music generation algorithms. The best results are set in \textbf{bold}. ↑: Higher is better. ↓: Lower is better. Content: Content Preservation. M\_Style: Musical Instrument Style Fit. S\_Style: Sound FX Style Fit. Quality: Overall Quality. Time: Inference Time (sec/sample). Memory: GPU Memory (GB).}
\resizebox{0.47\textwidth}{!}{
\begin{tabular}{cc|cc|c}
\hline
& & MusicGen \cite{copet2024simple} & MusicTI \cite{li2024music} & Stylus  \\ 
& & (NeurIPS'23) & (AAAI'24) & (Ours) \\ \hline \hline

\multirow{3}{*}{\makecell{Objective\\Evaluation ($\uparrow$)}}
 & Content  & 0.31 & 0.40 & \textbf{0.53} \\
 & M\_Style & 0.22 & \textbf{0.26} & 0.16 \\
 & S\_Style & 0.01 & 0.12 & \textbf{0.13} \\ \hline

\multirow{4}{*}{\makecell{User\\ Study ($\uparrow$)}}
 & Content  & 2.73 & 2.95 & \textbf{4.29} \\
 & M\_Style & 2.64 & \textbf{3.17} & \textbf{3.17} \\
 & S\_Style & 1.95 & 3.08 & \textbf{3.30} \\
 & Overall  & 2.67 & 2.69 & \textbf{3.38} \\ \hline

\multirow{2}{*}{Efficiency ($\downarrow$)}
 & Time   & 13.86 & 11.78 & \textbf{10.16} \\
 & Memory & 16    & \textbf{8.7}   & 18.1  \\ \hline

\end{tabular}}
\label{tab:quantitative}
\end{table}

%%%%%%%%%%%%%%%%%%%%%%%%%%%%%%%%%%%%%%%%%%%%%%%
%%%%%%%%%%%%%%%%%%%%%%%%%%%%%%%%%%%%%%%%%%%%%%%
% \multirow{3}{*}{\makecell{FAD ($\uparrow$)}}
%  & Content  & 8.57 & \textbf{9.15} & 8.69 \\
%  & M\_Style & 10.60 & 12.45 & \textbf{18.17} \\
%  & S\_Style & \textbf{20.66} & 13.46 & 18.46 \\ \hline

%  \multirow{3}{*}{\makecell{KL\\ Divergence ($\downarrow$)}}
%  & Content  & \textbf{0.88} & 1.02 & 0.69 \\
%  & M\_Style & \textbf{1.85} & 2.08 & 2.49 \\
%  & S\_Style & 4.23 & \textbf{3.46} & 4.18 \\ \hline

%  \multirow{3}{*}{\makecell{Chroma\\ Similarity ($\uparrow$)}}
%  & Content  & \textbf{0.68} & 0.61 & 0.62 \\
%  & M\_Style & \textbf{0.55} & 0.54 & 0.53 \\
%  & S\_Style & 0.52 & \textbf{0.53} & 0.52 \\ \hline
%%%%%%%%%%%%%%%%%%%%%%%%%%%%%%%%%%%%%%%%%%%%%%%
%%%%%%%%%%%%%%%%%%%%%%%%%%%%%%%%%%%%%%%%%%%%%%%

% ----Ablation v2----
\begin{table}[ht]
\renewcommand*{\arraystretch}{1.2}
\centering
\caption{\textbf{Ablation study} on style strength, content preservation, architectural components, and diffusion backbones. Higher values indicate better performance.}
{\footnotesize
\centering
\begin{tabularx}{0.48\textwidth}{cc|XXX}
\hline
& & Content & M\_Style & S\_Style \\ \hline \hline

& $\alpha$ = 0    & \textbf{0.720} & 0.153 & 0.096 \\
\multicolumn{1}{c|}{\multirow{2}{*}{Style}} & $\alpha$ = 0.25 & 0.678 & 0.154 & 0.101 \\
\multicolumn{1}{c|}{\multirow{2}{*}{Strength}} & $\alpha$ = 0.5  & 0.612 & 0.156 & 0.111 \\
\multicolumn{1}{c|}{} & $\alpha$ = 0.75 & 0.535 & \textbf{0.157} & 0.128 \\
& $\alpha$ = 1    & 0.454 & 0.156 & \textbf{0.146} \\ \hline

& $\gamma$ = 0    & 0.242 & 0.144 & 0.161 \\
\multicolumn{1}{c|}{\multirow{2}{*}{Content}} & $\gamma$ = 0.25 & 0.353 & 0.156 & 0.158 \\
\multicolumn{1}{c|}{\multirow{2}{*}{Preservation}} & $\gamma$ = 0.5  & 0.435 & 0.157 & 0.149 \\
\multicolumn{1}{c|}{} & $\gamma$ = 0.75 & 0.486 & \textbf{0.157} & \textbf{0.139} \\
& $\gamma$ = 1    & \textbf{0.519} & \textbf{0.158} & 0.133 \\ \hline \hline

\noalign{\vskip -7pt}\multicolumn{2}{c|}{} \\
\multicolumn{1}{c|}{LDM} & SDXL                          & 0.122 & 0.013 & -0.006 \\
\multicolumn{1}{c|}{Backbone} & {\textbf{SD v1.5 (Ours)}}     & \textbf{0.460} & \textbf{0.157} & \textbf{0.139} \\
\noalign{\vskip -7pt}\multicolumn{2}{c|}{}\\ \hline 

\multicolumn{2}{l|}{\textbf{Stylus (Ours)}} & \textbf{0.485} & \textbf{0.157} & \textbf{0.139} \\ 
\multicolumn{2}{l|}{\; w/o Guidance Scaling ($\alpha$)}           & 0.454 & 0.156 & \textbf{0.146} \\
\multicolumn{2}{l|}{\; w/o Init. AdaIN}                   & \textbf{0.520} & \textbf{0.161} & 0.133 \\

\multicolumn{2}{l|}{\; w/o Key features injection}       & 0.209 & 0.088 & 0.134 \\
\multicolumn{2}{l|}{\; w/o Value features injection}     & 0.152 & 0.082 & 0.117 \\ \hline

\multicolumn{2}{l|}{\textbf{Stylus (Ours)}} &  &  &  \\
\multicolumn{2}{l|}{\; w/ Phase-Preserving Reconst.} & \textbf{0.485} & 0.157 & \textbf{0.139} \\
\multicolumn{2}{l|}{\; w/ Griffin-Lim reconstruction} & 0.468 & \textbf{0.163} & 0.126 \\ \hline

\end{tabularx}}
\label{tab:ablation1}
\end{table}

\begin{table}[t]
\renewcommand*{\arraystretch}{1.1}
\centering
\caption{\textbf{Ablation study} of Multi-style Interpolation. $\beta \in [0,1]$ controls the relative contribution of two selected style references, Style A and Style B, smoothly transitioning from one to the other. Style A and B were chosen from a predefined set of references \cite{li2024music}.}
\small
\resizebox{0.47\textwidth}{!}{
\begin{tabular}{c|c|c}
\hline
& Style Fit w/ Style A ($\uparrow$) & Style Fit w/ Style B ($\uparrow$) \\ \hline \hline
$\beta$ = 0.1 & \textbf{0.153} & 0.100 \\
$\beta$ = 0.3 & 0.150 & 0.109 \\
$\beta$ = 0.5 & 0.145 & 0.120 \\
$\beta$ = 0.7 & 0.141 & 0.133 \\
$\beta$ = 0.9 & 0.137 & \textbf{0.146} \\ \hline
\end{tabular}}
\label{tab:ablation2}
\end{table}

\textbf{User Study} 
%To evaluate perceptual quality, we conducted a user study based on three criteria rated on a 5-point Likert scale (1 = lowest, 5 = highest). Participants evaluated (i) \textit{content preservation}, defined as the consistency of content sources' musical structure, (ii) \textit{style fit}, defined as the alignment of timbre and sound units with the reference style, and (iii) \textit{overall quality}, reflecting the naturalness and coherence of content–style fusion. Each participant was presented with 15 evaluation sets, each consisting of one content audio, one style audio, and three stylized outputs from MusicGen, MusicTI, and Stylus in randomized order. No time limits were imposed, allowing participants to freely compare and rate the outputs. Filtering incomplete submissions left 65 valid participants (out of 89), yielding 2,925 ratings that evaluated a single stylized output per content-style pair. As summarized in Table~\ref{tab:quantitative}, Stylus consistently achieved the highest scores across all evaluation criteria. In particular, the content preservation score is 57.1\% higher than MusicGen and 45.4\% higher than MusicTI, while still attaining substantially stronger style scores (M\_Style and S\_Style), indicating that our method effectively maintains the source content while transferring style. Moreover, the overall perceptual quality score of Stylus surpasses MusicGen by 26.6\% and MusicTI by 25.7\%, further demonstrating the strength of our attention-based design and phase-preserving reconstruction strategy. 
To evaluate perceptual quality, we conducted a 5-point Likert scale study. Participants evaluated (i) \textit{content preservation}, defined as the consistency of content sources' musical structure, (ii) \textit{style fit}, defined as the alignment of timbre and sound units with the reference style, and (iii) \textit{overall quality}, reflecting the naturalness and coherence of content–style fusion. From an initial 89 participants, 65 valid subjects remained after filtering, yielding 2,925 untimed ratings. Participants evaluated 15 sets, freely comparing a content-style audio pair alongside three randomized outputs from MusicGen, MusicTI, and Stylus. As summarized in Table~\ref{tab:quantitative}, Stylus consistently secured the highest scores across all evaluation criteria, significantly outperforming the baselines ($p < 0.05$, non-overlapping CIs). Most notably, our method yielded substantial improvements of up to 57.1\% in content preservation ($4.29 \pm 0.08$ vs.\ $\le 2.95$) and 26.6\% in overall perceptual quality ($3.38 \pm 0.08$ vs.\ $\le 2.69$). Furthermore, Stylus demonstrated a clear superiority in Sound FX style fit ($3.30 \pm 0.12$) while achieving parity with MusicTI in musical instrument styles. Collectively, these results provide compelling validation for our attention-based manipulation and phase-preserving reconstruction strategies.

%%%%%%%%%%%%%%%%%%%%%%%%
\subsection{Ablation Study}
We conducted ablation experiments to justify our design choices and examine the role of key components. Our analysis covers three aspects: (i) \textbf{Stylization Manipulation}, where $\alpha$ modulates style strength, while $\gamma$ regulates preservation of the content structure; (ii) \textbf{Architectural Components}, including the impact of guidance scaling ($\alpha$), the initial AdaIN layer, phase-preserving reconstruction, key/value injection, and the choice of LDM backbone; and (iii) \textbf{Multi-style Interpolation}, where linear mixing between two style references is controlled by $\beta$.

Table~\ref{tab:ablation1} quantifies the contribution of each component. Increasing $\alpha$ enhances style adherence at the cost of content preservation, confirming an inherent trade-off. A higher query preservation weight ($\gamma$) benefits structured instrument styles by maintaining articulation, whereas it diminishes performance on texture-heavy sound effects that require stronger style dominance. The ablation also validates that guidance scaling ($\alpha$) is critical for capturing temporal structure in instruments, while both AdaIN and phase preservation are important for preserving event-based textures in sound effects. Notably, dropping either key or value injections severely impacts all metrics, underscoring their combined role. Among backbone models, \textit{SD v1.5} yields the best performance. Crucially, beyond its impact on specific textures, phase-preserving reconstruction provides a robust temporal scaffold. Although it may trade off subtle phase-dependent stylistic cues, it stabilizes rhythm and onsets, yielding outputs with substantially higher perceptual quality and musical coherence than those produced by iterative phase estimation methods. Finally, as shown in Table~\ref{tab:ablation2}, varying $\beta$ enables smooth, continuous transitions between styles, offering a lightweight mechanism for multi-style blending. Our default STFT configuration ($n_{\text{fft}} = 17,640$) ensures optimal 2D structural clarity for content preservation (Content: $0.5023$, M\_Style: $0.1724$, S\_Style: $0.0843$), preventing the severe structural degradation (Content: $0.2072$) and artificial stylistic overfitting (M\_Style: $0.2068$, S\_Style: $0.1269$) caused by halving the window size ($n_{\text{fft}} = 8,820$).

\section{Conclusion}
\label{sec:conclusion}

In this work, we presented Stylus, a training-free framework that repurposes pretrained image diffusion models for zero-shot music style transfer. By treating Mel-spectrograms as structured images, Stylus leverages attention-based key-value injection and phase-preserving reconstruction to achieve a superior balance between stylistic fidelity and structural preservation. While relying on generic spatial priors enables robust manipulation without task-specific training, it inherently trades off sensitivity to fine-grained acoustic cues, such as phase dependencies. Future work will focus on bridging this gap, aiming to enhance acoustic nuance recovery while retaining the flexibility of our training-free inference.

\vfill\pagebreak

\bibliographystyle{IEEEbib}
\bibliography{strings,refs}

\end{document}